\documentclass[aps,prl,reprint]{revtex4-1}
	\usepackage{setspace}
	\usepackage{amsmath}    
	\usepackage{graphicx}   
	\usepackage{verbatim}   
	\usepackage{color}      
	\usepackage{subfigure}  
	\usepackage{hyperref}   
	\usepackage{dcolumn}   
\makeindex

\begin{document}

\pagestyle{myheadings}

\title{Comparison of spectral linewidths for quantum degenerate bosons and fermions}	

\author{R.P.M.J.W. \surname{Notermans}}
\author{R.J. \surname{Rengelink}}
\author{W. \surname{Vassen}}
\email{w.vassen@vu.nl}
\affiliation{LaserLaB, Department of Physics and Astronomy, Vrije Universiteit, De Boelelaan 1081, 1081 HV Amsterdam, Netherlands}

\begin{abstract}
We observe a dramatic difference in optical line shapes of a $^4\text{He}$ Bose-Einstein condensate and a $^3\text{He}$ degenerate Fermi gas by measuring the 1557-nm $2~^3S-2~^1S$ magnetic dipole transition ($8~\text{Hz}$ natural linewidth) in an optical dipole trap. The $15~\text{kHz}$ FWHM condensate line shape is only broadened by mean field interactions, whereas the degenerate Fermi gas line shape is broadened to $75~\text{kHz}$ FWHM due to the effect of Pauli exclusion on the spatial and momentum distributions. The asymmetric optical line shapes are observed in excellent agreement with line shape models for the quantum degenerate gases. For $^4$He a triplet-singlet s-wave scattering length $a=+50(10)_{\text{stat}}(43)_{\text{syst}}~a_0$ is extracted. The high spectral resolution reveals a doublet in the absorption spectrum of the BEC, and this effect is understood by the presence of a weak optical lattice in which a degeneracy of the lattice recoil and the spectroscopy photon recoil leads to Bragg-like scattering.
\end{abstract}


\maketitle

The bosonic or fermionic nature of a particle is a fundamental property, and trapped quantum degenerate gases display dramatic different behaviour depending on the quantum statistical nature of the gas. At low temperatures identical bosons accumulate in the lowest state in the trap, leading to Bose-Einstein condensation. In contrast, identical fermions cannot occupy the same state due to the Pauli exclusion principle, and will `fill' all states in the trap from the bottom up until no more atoms - or states - are available. A drastic difference in line shape of a narrow optical transition is expected when measured in a Bose-Einstein condensate (BEC) and a degenerate Fermi gas (DFG). In this Letter we show a direct comparison of this difference between a BEC of metastable $^4\text{He}$ and a DFG of metastable $^3\text{He}$ trapped in an optical dipole trap (ODT).

We do this work in the framework of high-precision frequency metrology in helium, aimed at testing quantum electrodynamics (QED). Comparison of accurate transition frequencies is used to determine fundamental physical parameters that are difficult to measure otherwise, such as the nuclear charge radius of an atom. Recently high-precision frequency metrology in (muonic) hydrogen and deuterium resulted in a remarkable discrepancy in the determination of the proton and deuteron charge radius \cite{Antognini2,Pohl1}. This discrepancy, also known as the `proton radius puzzle', is currently under scrutiny by many groups all over the world and similar work is ongoing for helium \cite{Antognini1}. To determine the  $^3\text{He}$-$^4\text{He}$ nuclear charge radius difference, we recently measured the doubly forbidden $2~^3S-2~^1S$ transition at $1557~\text{nm}$ (natural linewidth 8 Hz) in both quantum degenerate  $^4\text{He}$ and  $^3\text{He}$ with 1.8 kHz and 1.5 kHz accuracy, respectively \cite{Rooij1}. The measured isotope shift, combined with QED calculations, allowed a determination of a squared nuclear charge radius difference of $1.028(11)~\text{fm}^2$ \cite{Pachucki1}. To compare this determination to measurements in muonic helium ions \cite{Antognini1} we aim to measure the $2~^3S-2~^1S$ transition frequency with $\ll 1$ kHz accuracy. Using a narrow linewidth spectroscopy laser we are able to observe asymmetric line shapes for a BEC and a DFG of metastable helium as well as a line splitting in the optical spectrum of the BEC. Quantification of these effects by understanding the line shapes is essential in achieving the sub-kHz accuracy goal.

Our experimental setup is similar to earlier work \cite{Rooij1} and to a more recent measurement of the $2~^3S_1-2~^1P_1$ transition at 887 nm \cite{Notermans1}. We load a BEC of typically $10^6$ atoms in the metastable $2~^3S_1\ (m_J=+1)$ state (lifetime $\sim 7800$ s, internal energy $19.82$ eV) into a crossed-beam ODT operating at 1557.3 nm. The crossing angle between the ODT beams is $19^{\circ}$, and the temperature of the thermal atoms in the ODT is typically $T\approx 0.2~\mu$K. As the fermionic $^3\text{He}$ atoms cannot thermalize once their temperature is below the p-wave barrier, they are loaded simultaneously with $^4\text{He}$ and sympathetically cooled to degeneracy \cite{McNamara1}. The quantum degenerate $^3\text{He}$-$^4\text{He}$ mixture is loaded into the ODT to rethermalize, after which the $^4\text{He}$ atoms are blown away using resonant light. This procedure leaves a pure DFG of thermalized $^3\text{He}$ in the $2~^3S_1~(F=\frac{3}{2},m_F=+\frac{3}{2})$ state. The spectroscopy beam copropagates with one of the ODT beams in order to overlap with the trapped cloud. The atoms are probed for a few seconds, after which the remaining cloud is released from the ODT. The time-of-flight signal of the metastable atoms is measured on a microchannel plate (MCP) detector and used to determine the remaining atom number, temperature and chemical potential. The measurements alternate with and without the spectroscopy light in order to have a continuous background measurement to normalize the line shapes. 

For this experiment a narrow linewidth fiber laser is transfer-locked to an ultrastable ($<2~\text{Hz}$) laser system operating at $1542~\text{nm}$ using a caesium clock-referenced femtosecond frequency comb to bridge the $15~\text{nm}$ wavelength difference between both lasers. Due to uncompensated fiber links we estimate a residual $\sim 5~\text{kHz}$ linewidth of the spectroscopy laser, which is in agreement with the $4.5~\text{kHz}$ linewidth (FWHM) determined in our line shape fits. This is a factor 20 improvement compared to our previous experiment \cite{Rooij1}. 

\begin{figure}[t]
	\begin{center}
		\includegraphics[width=0.85\columnwidth]{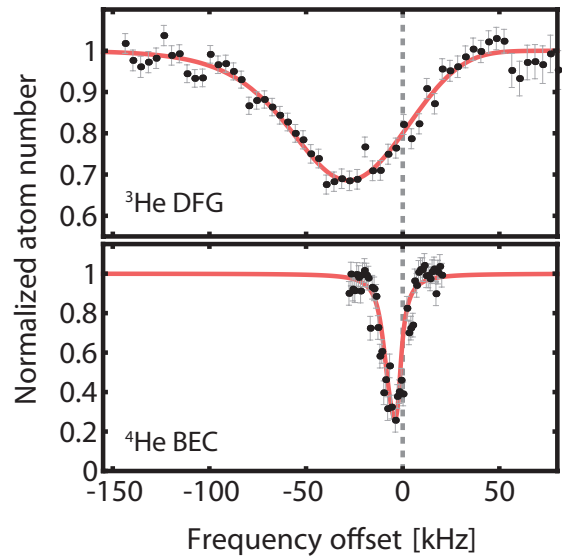}
		\caption{Direct comparison of the (normalized) optical line shapes of the $2~^3S-2~^1S$ transition measured in a degenerate Fermi gas (top) and a Bose-Einstein condensate (bottom) of metastable helium. The full lines represent the fits provided by the models discussed in the main text, and display a small but significant asymmetry. For a clear comparison only one peak of the observed BEC doublet is shown (see Fig. \ref{fig:becdepletion}). The zero frequency represents the transition frequency from the bottom of the trap.}
		\label{fig:becanddfgline}
	\end{center}
\end{figure}

Fig. \ref{fig:becanddfgline} shows the optical $\sigma^{-}$ transitions measured in a BEC [$2~^3S_1~(m_J=+1) \to 2~^1S_0~(m_J=0)$] and DFG [$2~^3S_1~(F=\frac{3}{2},m_F=+\frac{3}{2}) \to 2~^1S_0~(F=\frac{1}{2},m_F=+\frac{1}{2})$]. The uncertainty in the frequency is $1.8~\text{kHz}$, and the error bars in the normalized atom numbers are based on the atom number fluctuations in the measurements. The zero on the frequency axis represents the transition frequency from the bottom of the trap which is not measured as an absolute frequency. For the DFG results the atom number $N\approx~3~\times~10^5$ and peak density $\sim 1~\times~10^{12}~\text{cm}^{-3}$. There are three times as many atoms in the BEC compared to the DFG due to the more complicated loading procedure of the DFG \cite{McNamara1}, and the peak density of the BEC is ten times higher. Despite this, the line shape of the DFG is over five times broader. This is caused entirely by the broad momentum and spatial distribution of the fermions. In contrast, the BEC line shape only has a finite width due to the mean field interactions (which are absent in a coherent excitation of a Fermi gas \cite{Zwierlein1}) and the linewidth of the spectroscopy laser. Without the effects of quantum statistics the width of both line shapes would simply be the Doppler width ($31~\text{kHz}$ for $^4\text{He}$, $35~\text{kHz}$ for $^3\text{He}$). This huge difference in linewidths based on the quantum statistics of the helium isotopes is complementary to the observation of bunching and antibunching with the same atoms \cite{Jeltes1}. For frequency metrology purposes it is clear that proper modeling is imperative in order to determine the true transition frequency.

The line shape for the DFG is calculated using the absorption line profile from Ref. \cite{Juzeliunas1} and involves explicit integration of the Fermi-Dirac distribution of the spatial and momentum states occupied in the ODT, convolved with a Lorentzian distribution with a FWHM of $4.5~\text{kHz}$ (determined from the BEC fits) to model the finite linewidth of the spectroscopy laser. Time-dependent depletion of the DFG does not play a role because the fermions neither rethermalize nor redistribute over the trap states during the optical excitation. Using the experimentally determined degeneracy $T/T_F~=~0.33(7)$ and chemical potential $\mu=0.55(15)~\mu\text{K}$ of the DFG, the calculated line shape is shown in Fig.~\ref{fig:becanddfgline}~(top). As only the relative amplitude and frequency offset of the line are fitted to the data, the model predicts the line shape perfectly. Although hardly visible, the line shape is asymmetric and the model provides a reduced $\chi^2=1.09$.

The line shape for light absorption from a BEC is fundamentally different from that of a DFG and was first calculated by Killian for the absorption on the $1S-2S$ transition in a hydrogen BEC \cite{Fried1,Killian1}. Excellent agreement with the data was demonstrated, but the line shape function \cite{Killian2} cannot be used in our experiment for two reasons. First, in \cite{Killian2} it is assumed that the trapping potentials of both the initial and final state are equal. This assumption is invalid in our ODT as the ratio of the polarizabilities of both states $\alpha_{s}/\alpha_{t}=-1.64(1)$ \cite{Notermans2}, where $s$ and $t$ denote the singlet and triplet state ($2~^1S$ atoms are repelled from the trap). Second, the excitation fraction in \cite{Killian2} was on the order of $1\%$ and therefore depletion of the condensate during excitation could be neglected. This is invalid in our experiment as the excited BEC fraction is typically $20-70\%$ to have an acceptable signal-to-noise ratio. Therefore we extend the Killian model \cite{Killian2} by including the polarizabilities in the effective potentials of the initial and final state \cite{Supplemental}. This results in the addition of the ac Stark shift to the resonance condition, and an effective rescaling of the mean field shift term $(4\pi\hbar^2n_0/m)(a_{ts}-a_{tt})$ which becomes $(4\pi\hbar^2n_0/m)(a_{ts}-(\alpha_s/\alpha_t)a_{tt})$. Here $n_0$ is the peak density of the condensate, $a_{tt}$ the $2~^3S_1-2~^3S_1$ s-wave scattering length in the pure $^5\Sigma^+_g$ potential, and $a_{ts}$ the $2\ ^3S_1(m_J=+1)-2\ ^1S_0(m_J=0)$ s-wave scattering length. Although $a_{ts}$ has not been measured or calculated to date, $a_{tt}$ is very accurately known: $a^{\text{theory}}_{tt}~=~143.0(5)~a_0$ \cite{Przybytek2} and $a^{\text{exp}}_{tt}~=~142.0(1)~a_0$ \cite{Moal1}, where $a_0$ is the Bohr radius.

It is convenient to express the line shape of the BEC using the chemical potential $\mu=4\pi\hbar^2a_{tt}n_0/m$, which we determine directly from a time-of-flight measurement. The line shape $S(\nu,\mu)$ is \cite{Supplemental}
\begin{align}
S(\nu,\mu) = \frac{15 \pi \hbar \Omega_R^2}{4} N \frac{h \nu}{\tilde{\mu}^2} \sqrt{1 + \frac{h\nu}{\tilde{\mu}}}, \label{eqn:lineshape}
\end{align}
where $\Omega_R$ is the Rabi frequency of the transition, $N$ the total atom number of the BEC, $\nu$ the detuning from the absolute transition frequency including the full ac Stark shift of the trap, and $\tilde{\mu}=(a_{ts}/a_{tt}-\alpha_{s}/\alpha_{t})\mu$ the rescaled chemical potential of the BEC. This rescaling shows how the mean field interaction and ac Stark shift affect the effective potential experienced by the atoms. The line shape of the BEC is asymmetric with a high-frequency cut-off at $\nu = 0$. 

As the atom number of the condensate scales as $N \propto \mu^{5/2}$ in the Thomas-Fermi limit and the line shape $S(\nu,\mu)$ constitutes a one-body loss process, the decay of the chemical potential of the BEC during the spectroscopy phase can be written as \cite{Supplemental}
\begin{align}
\frac{d\mu}{dt} = \frac{2}{5} \frac{\mu}{N} \tilde{S}(\nu,\mu) - \frac{2}{5} \Gamma \mu, \label{eqn:diffeqn}
\end{align}
where $\tilde{S}(\nu,\mu)$ is the line shape $S(\nu,\mu)$ convolved with a Lorentzian distribution to model the spectroscopy laser linewidth \cite{Supplemental}. We include the one-body lifetime $\Gamma^{-1}$ of the gas as the typical interaction times are long enough (1-6 seconds) that one-body loss cannot be neglected. The decay of the chemical potential is slow enough such that the condensate can be assumed to remain in equilibrium throughout the excitation \cite{Supplemental}. The BEC is held in the ODT for 4-5 seconds before switching on the probe light so two- and three-body loss processes are negligible. The nonlinear differential Eq. \ref{eqn:diffeqn} is numerically solved to fit to the line shape as shown in Fig. \ref{fig:becanddfgline} (bottom). Here we use only the frequency offset and $a_{ts}$ scattering length as free parameters, giving a reduced $\chi^2=0.94$.

\begin{figure}[t]
	\begin{center}	
		\includegraphics[width=0.85\columnwidth]{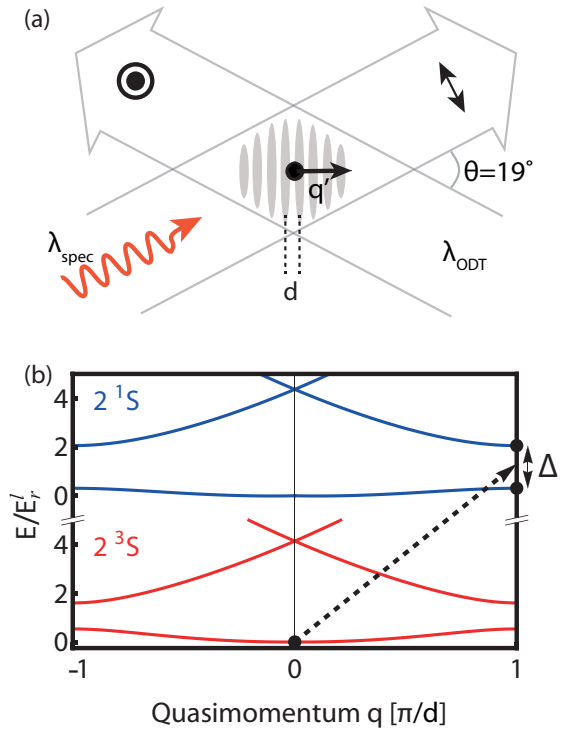}
		\caption{(a) In the crossed-beam optical dipole trap geometry we have a weak optical lattice with periodicity $d~=~\lambda_{\text{ODT}}/[2 \cos(\theta/2)]$. Absorption of a spectroscopy photon results in a recoil momentum $q'~=~2\pi \cos(\theta/2)/\lambda_{\text{spec}}$ in the lattice direction equal to the lattice recoil momentum $\pi/d$. (b) Band structure of the optical lattice for the ground (lower red bands) and upper state (upper blue bands) for a typical lattice amplitude $V_0\approx 2E^l_r$. The BEC is situated at quasimomentum $q=0$ in the lowest band (black dot). Absorption of a photon creates a quasimomentum $q'\approx \pi/d$ in the optical lattice for the excited state. Absorption can take place if the spectroscopy laser frequency is resonant with the lowest or first band at the edge of the Brillouin zone (black dots at $q=\pi/d$), giving rise to the observed bandgap splitting $\Delta$.}
		\label{fig:bandstructure}
	\end{center}
\end{figure}

\begin{figure}[t]
	\begin{center}
		\includegraphics[width=0.87\columnwidth]{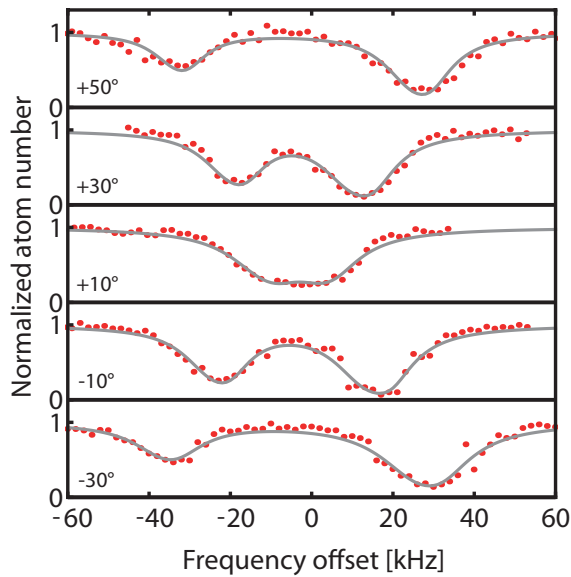}
		\caption{Absorption spectrum of a BEC in a weak optical lattice for various rotation angles of the polarization of the second ODT beam with respect to the first ODT beam from the configuration shown in Fig. \ref{fig:bandstructure}(a). The spectra are offset and centered around the midway frequency of the two lines, and the lines are fits of the time-dependent line shape model. The spectroscopy interaction times used in these measurements are (top to bottom): 1.5 s, 2 s, 1.5 s, 4 s, and 6 s and vary as the Rabi frequency also varies with the rotation angles.}
		\label{fig:linesplitting1}
	\end{center}
\end{figure}

Interestingly, we observe a doublet in the BEC spectrum where a single peak was expected. This double peak structure is attributed to the presence of a weak optical lattice in our crossed dipole trap due to birefringence in our vacuum windows. The ODT laser wavelength $\lambda_{\text{ODT}}\approx 1557.3~\text{nm}$ (sufficiently off-resonance from the $2~^3S-2~^1S$ transition to have negligible scattering) is close to the transition wavelength and creates a lattice with periodicity $d~=~\lambda_{\text{ODT}}/[2 \cos(\theta/2)]$ and effective lattice recoil energy $E^l_r~=~\hbar^2q_l^2/2m$, where $q_l=\pi/d$. This recoil energy is nearly degenerate with the recoil when absorbing a spectroscopy photon in the lattice frame, $E_r~=~\hbar^2q'^2/2m \approx h \times 20.7~\text{kHz}$, with $q'=2\pi \cos(\theta/2)/\lambda_{\text{spec}}$, see Fig. \ref{fig:bandstructure}(a). The absorbed spectroscopy photon provides the excited wavefunction a quasimomentum $q'$ in the frame of the lattice. This quasimomentum is at the edge of the first Brillouin zone and therefore at the optical lattice bandgap, as shown in Fig. \ref{fig:bandstructure}(b). The resonance condition can only be satisfied below or above the bandgap, leading to a line splitting $\Delta$ of the transition, where $\Delta~=~V_0/2$ and $V_0$ is the optical lattice modulation amplitude as observed by the excited state \cite{Morsch1,Bloch1}. 

This excitation in a weak optical lattice is reminiscient of Bragg scattering of a BEC in an applied optical lattice \cite{Kozuma1,Ernst1}. Contrary to Bragg scattering, where an applied moving optical lattice causes diffraction, the direct one-photon optical excitation causes the transition to a higher momentum state near the edge of the Brillouin zone. We verify the presence of the weak optical lattice by rotating the polarization of the second ODT beam with respect to the first. Fig. \ref{fig:linesplitting1} shows that the splitting increases as $V_0$ is increased and in these measurements we estimate the optical lattice modulation amplitude for the $2~^1S$ state to be $V_0\leq~6.5E^l_r$ for the largest splitting shown. As the polarizability for the $2~^3S$ atoms is smaller by a factor $1.64$, the optical lattice observed by the BEC is $V_0\leq~4.0E^l_r$ for the largest splitting and the ultracold cloud is in the superfluid regime \cite{Morsch2}. Aspect ratio inversion in absorption images of the expanding cloud confirms this. In this regime the mean field description is applicable and coupling to higher lattice bands can be ignored. The doublet is simultaneously fit with the same model and fixed experimental parameters, apart from the line splitting and amplitude ratio, as shown in Fig. \ref{fig:linesplitting1}. For the DFG line shape measurements we have minimized $V_0$ by looking at the BEC spectra shown in Fig. \ref{fig:linesplitting1}. At this setting the DFG line shape is much broader than the effect of the lattice or, equivalently, the Fermi energy $E_F \gg V_0$. 

\begin{figure}[t]
	\begin{center}
		\includegraphics[width=0.85\columnwidth]{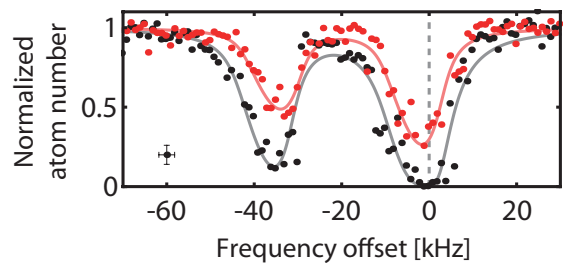}
		\caption{Double-peak structure of the Bose-Einstein condensate absorption spectrum due to the weak optical lattice measured for a spectroscopy laser probe time of 1 s (top, red) and 3 s (bottom, black). The uncertainty per data point is indicated by the bottom left inset. The full lines are fits of the time-dependent line shape model. For the top (red) and bottom (black) fit we find $\chi^2=0.9$ and $\chi^2=1.3$, respectively. A single absorption line from the top (red) dataset is used in Fig. \ref{fig:becanddfgline}.}
		\label{fig:becdepletion}
	\end{center}
\end{figure}

We measure the time-dependent behaviour of the BEC line shapes to extract the scattering length $a_{ts}$, which is the only unknown parameter in the line shape calculations. The optical lattice operates with splitting $\Delta \approx 35$ kHz such that the lattice is as weak ($V_0~\approx~2E^l_r$) as possible but the two lines are separated sufficiently so they can be individually resolved. Background and lifetime measurements provide the one-body loss rate $\Gamma^{-1} \approx 10$ s and the chemical potential of the BEC at $t=0$. The scattering length $a_{ts}$ is determined by simultaneous fitting of six doublet lines with interaction times ranging from 0.5 s to 3 s, and Fig. \ref{fig:becdepletion} shows the lines for 1 s and 3 s. The average reduced $\chi^2$ of all fits is $1.1$, showing good agreement of the model with the data. From the fits we find $a_{ts}~=~+50(10)_{\text{stat}}(43)_{\text{syst}}~a_0$. The statistical uncertainty is a $1\sigma$ uncertainty based on simultaneous $\chi^2$ minimization of all data sets. The systematic uncertainty is a worst-case error bound based on our estimation of the Rabi frequency $\Omega_R~=~2\pi\times~21(5)$ Hz \cite{Supplemental}. Our result is in agreement with the estimated range of possible scattering lengths based on previous mean field shift measurements \cite{Rooij1}. Furthermore the determination is in agreement with a surprisingly accurate theoretical value $a_{ts}~=~+42.5^{+0.5}_{-2.5}~a_0$~\cite{Cocks1}, based on \textit{ab initio} $1~^3\Sigma_g^+$ and $2~^3\Sigma_g^+$ molecular potentials \cite{Mueller1} including large ionization widths which make the calculations insensitive to the actual coupling between the potentials.

To conclude, we have directly compared the fundamental difference between quantum degenerate fermions and bosons by measuring and calculating the asymmetric absorption line shapes of a Bose-Einstein condensate and a degenerate Fermi gas of metastable helium. The line shape of the Fermi gas shows excellent agreement without any adaptations to the existing model \cite{Juzeliunas1}. We extended the line shape of the Bose-Einstein condensate from the existing model \cite{Killian2} to include ac Stark shift and time-dependent depletion of the condensate. The model shows good agreement with the data, and the $2~^3S-2~^1S$ s-wave scattering length is extracted to be $a_{ts}=+50(10)_{\text{stat}}(43)_{\text{syst}}~a_0$, in good agreement with scattering length calculations.

We also show how a weak optical lattice can induce a line splitting if the lattice recoil is degenerate with the spectroscopy photon recoil. The effect is similar to Bragg scattering and allows observation of the lattice in the optically excited state. Measurement of the line splitting and the total ac Stark shift on the transition frequency would allow determination of both the dynamic polarizability of the ground and excited states. Furthermore, if unresolved, this effect could lead to a frequency broadening or shift in any spectroscopy measurement in an optical dipole trap.

\acknowledgments
We gratefully acknowledge K.S.E. Eikema for providing us the use of the ultrastable laser and frequency comb and R. Kortekaas for excellent technical support. We are also grateful to D. Cocks and I. Whittingham for calculating the $2\ ^3S-2\ ^1S$ s-wave scattering length. This work is part of the research programme of the Foundation for Fundamental Research on Matter (FOM), which is financially supported by the Netherlands Organization for Scientific Research (NWO).

\bibliography{paper_references}
\bibliographystyle{apsrev4-1}

\newpage

\onecolumngrid
\appendix
\section{Supplemental material}
In this supplemental material to our paper `Comparison of spectral linewidths for quantum degenerate bosons and fermions' we provide the full derivation of the time-dependent line shape model of optical absorption in a Bose-Einstein condensate, and our estimation of the Rabi frequency used in the line shape fits.

\section{BEC line shape model}
Using the line shape model as calculated by Killian \cite{Fried1,Killian1} as a starting point, we introduce the ac Stark shift and the time-dependent behaviour in two steps. Finally we give the analytic solution of the line shape convolved with a Lorentzian laser line shape.

\subsection{Adding an ac Stark shift}
Although the hydrogen experiment is done in a magnetic trap \cite{Killian2}, this does not influence the mathematics for the optical dipole trap (ODT) case. The trapping potential of an ODT is proportional to the dynamic polarizability $\alpha$ and the local intensity $I(\vec{r})$ of the optical field as $V(\vec{r}) \propto -\alpha I(\vec{r})$. As $I(\vec{r})$ is identical for both atomic states, we can express the potential of the excited state simply as $V_2(\vec{r}) = (\alpha_2/\alpha_1) V_1(\vec{r})$, where $\alpha_1$ and $\alpha_2$ are the dynamic polarizabilities of the ground and excited states, respectively. In the main paper the ground and excited states are labeled $t$ and $s$ respectively to distinguish between the singlet and triplet states.

The resonance condition of Killian includes a spatially dependent mean field shift and we add another spatially dependent term here. Using the general resonance condition
\begin{align}
h(\nu-\nu_0) = V_2^{\text{eff}}(\vec{r}) - V_1^{\text{eff}}(\vec{r}),
\end{align}
where $V_{1,2}^{\text{eff}}(\vec{r})$ are the effective potentials of the ground and excited state. Note that the definition of $\nu$ is slightly different from the definition in Eq. 2 in the main paper as we have omitted the ac Stark, Zeeman and recoil shift from the definition for simplicity of notation. Filling in the mean field interaction and the ac Stark shift, we get
\begin{align}
h(\nu-\nu_0) = \Big( \frac{\alpha_2}{\alpha_1} - 1 \Big) V_1(\vec{r}) + \Delta U n_1(\vec{r}). \label{eqn:firstresonance}
\end{align}
Here $\Delta U n_1(\vec{r}) = (U_2 - U_1) n_1(\vec{r})  = (4 \pi \hbar^2 n_1(\vec{r}) ) / m) (a_{21}-a_{11})$ is the mean field interaction term, with atomic mass $m$, and scattering lengths $a_{21}$ and $a_{11}$ for the excited-ground state and ground-ground state collisions, respectively. The ground state density profile is given by $n_1(\vec{r})$. We see how inclusion of the ac Stark effect leads to a spatial dependence of the resonance condition similar to the mean field interactions.

Mean field interactions between two $2~^1S_0$ atoms are neglected as the estimated fraction of $2~^1S_0$ atoms is limited to a worst-case upper bound of $0.2\%$ in the measurement. This estimate is based on the fact that the $2~^1S_0$ atoms experience an antitrapping potential and are therefore expelled from the trapping region at a timescale shorter than 2 ms. We note that this timescale is an upper limit based on worst-case estimates of atoms leaving the trap from the center along the long axial direction. In most cases the timescale is much shorter and on the order to $10~\mu\text{s}$, which makes the excited state fraction even lower. As the radiative lifetime of the $2~^1S_0$ state is $20~\text{ms}$, spontaneous decay does not play a role in this process.

We can use the linear relationship between the ground state trapping potential and the density distribution of the BEC through the Thomas-Fermi relation $\mu - U_0 = U_1 n_1(\vec{r}) + V_1(\vec{r})$, where $U_0 = V_1(r = 0)$ is the depth of the trap. This definition is slightly unconventional, and is caused by the fact that most traps are defined as $V(r = 0) = 0$ for convenience. In our case we have defined $V_1(r \to \infty) = 0$, which leads to the slight modification. Substituting this in Eq. \ref{eqn:firstresonance}, we find
\begin{align}
h(\nu - \nu_0) = \Big( \frac{\alpha_2}{\alpha_1} - 1 \Big) (\mu - U_0) + \Big( U_2 - \frac{\alpha_2}{\alpha_1} U_1 \Big) n_1(\vec{r}). \label{eqn:acstarkresonance}
\end{align}
This resonance condition now only depends on the spatial dependence of the density distribution of the BEC, and any further analysis is completely analogous to the work done by Killian. An important check is to see what happens if the trapping potentials are equal, which is the case for the hydrogen experiment or, indeed, a magic wavelength ODT. In this case $\alpha_1 = \alpha_2$ and Eq. \ref{eqn:acstarkresonance} indeed reduces to the resonance condition as derived by Killian.

\subsection{New lineshape}
Using the definition by Killian to calculate the line shape for a Doppler-sensitive profile in a spherically symmetric trap, we start with
\begin{align}
S(\nu) = \pi \hbar \Omega_R^2 \int 4 \pi r^2 n_1 (r) \text{d} r \delta \Big[ h (\nu - \nu_0) + \Big( \frac{\alpha_2}{\alpha_1}-1 \Big) (U_0 - \mu) - \Big( U_2 - \frac{\alpha_2}{\alpha_1} U_1 \Big) n_1 (r)  \Big],
\end{align}
where $\Omega_R$ is the Rabi frequency. This integral has the analytical solution
\begin{align}
S(\nu) = \frac{15 \pi \hbar \Omega_R^2 N}{4} \frac{h (\nu - \nu_0) - \Big( \frac{\alpha_2}{\alpha_1} -1 \Big) (U_0 - \mu)}{\Big( \frac{U_2}{U_1} - \frac{\alpha_2}{\alpha_1} \Big)^2 \mu^2}
\times \sqrt{1 + \frac{h (\nu - \nu_0) - \Big( \frac{\alpha_2}{\alpha_1} -1 \Big) (U_0 - \mu)}{\Big( \frac{U_2}{U_1} - \frac{\alpha_2}{\alpha_1} \Big) \mu}}, \label{eqn:acstarklineshape}
\end{align}
with $N$ the total number of atoms in the condensate. Eq. \ref{eqn:acstarklineshape} is also given in the main text as Eq. 2. This function is valid in the frequency domain
\begin{align}
\Big( \frac{\alpha_2}{\alpha_1} -1 \Big) (\mu - U_0) \leq h (\nu - \nu_0) \leq \Big( \frac{\alpha_2}{\alpha_1} -1 \Big) (\mu - U_0) + \Big( \frac{U_2}{U_1} - \frac{\alpha_2}{\alpha_1} \Big) \mu.
\end{align}
This looks quite similar to Eq. 25 from Killian \cite{Killian2}, which becomes more apparent if the ac Stark effect is removed by setting $\alpha_1 = \alpha_2$ i.e. the `magic wavelength' condition. Also note that $U_2/U_1 = a_{21}/a_{11}$ and therefore this result has the nice feature of being dependent on the - dimensionless - ratios of the polarizabilities and scattering lengths. 

\subsection{Depletion of the condensate during excitation}
The weak excitation approximation - as used before - assumes that the chemical potential of the condensate does not change significantly during the excitation. In the hydrogen BEC work this is a good approximation as only a fraction of $10^{-2}$ of the condensate is excited. In our case the excited fraction reaches over $50\%$ and the change of the chemical potential during the excitation has to be taken into account in the line shape calculations.

In the Thomas-Fermi limit the relationship between the chemical potential $\mu$ and the atom number $N$ of a BEC is the well-known nonlinear relationship
\begin{align}
N = \frac{2^{\frac{5}{2}}}{15 \sqrt{m} \hbar^2 a \bar{\omega}^3} \mu^{\frac{5}{2}}, \label{eqn:TFrelation}
\end{align}
where $m$ is the atomic mass, $a$ the s-wave scattering length and $\bar{\omega} = (\omega_x \omega_y \omega_z)^{\frac{1}{3}}$ the geometric averaged trap frequency. The atom number loss and chemical potential loss are then related as
\begin{align}
\frac{dN}{dt} = \frac{dN}{d\mu} \frac{d\mu}{dt} = \frac{5}{2} \frac{2^{\frac{5}{2}}}{15 \sqrt{m} \hbar^2 a \bar{\omega}^3} \mu^{\frac{3}{2}} \frac{d\mu}{dt} = \frac{5}{2} \frac{N}{\mu} \frac{d\mu}{dt}. \label{eqn:TFdiffequation}
\end{align}
Realizing that the function $S(\nu)$ represents the one body atom number loss $dN/dt$, we can use this relationship to use Eq. \ref{eqn:acstarklineshape} as a differential equation for the chemical potential:
\begin{align}
\frac{d\mu}{dt} = \frac{2}{5} \frac{\mu}{N} S(\nu) =  
\frac{3 \pi \hbar \Omega_R^2}{2} \frac{h (\nu - \nu_0) - \Big( \frac{\alpha_2}{\alpha_1} -1 \Big) (U_0 - \mu(t))}{\Big( \frac{U_2}{U_1} - \frac{\alpha_2}{\alpha_1} \Big)^2 \mu} 
\times \sqrt{1 + \frac{h (\nu - \nu_0) - \Big( \frac{\alpha_2}{\alpha_1} -1 \Big) (U_0 - \mu(t))}{\Big( \frac{U_2}{U_1} - \frac{\alpha_2}{\alpha_1} \Big) \mu(t)}}.  \label{eqn:diffequation}
\end{align}
Solving this differential equation will result in a function $\mu(\nu,t)$ which shows how the chemical potential of the condensate changes as function of the laser frequency and interaction time. 

The above derivation assumes that the condensate stays in equilibrium throughout the excitation, although we can achieve significant excitation fractions in the experiment. To see if the condensate can redistribute fast enough to assume equilibrium, we can compare the change in density distribution to the sound velocity of the condensate. As the density distribution scales linearly with the chemical potential, the estimated upper limit on the relative change in the density distribution is
\begin{align}
\frac{1}{\mu} \frac{d\mu}{dt} = \frac{2}{5} \frac{1}{N} \frac{dN}{dt} \approx 0.13~\text{s}^{-1},
\end{align}
where we assume a (large) atom number of $N = 10^6$ atoms and a (short) excitation time of $3$ s based on the achieved depletion in Figure 4 in the Letter. The sound velocity is
\begin{align}
c = \sqrt{\frac{\mu}{m}} \approx 2~\text{cm/s},
\end{align}
which, for a condensate with a length of $\sim 200~\mu\text{m}$ (which is an upper limit), corresponds to a frequency of the density oscillations of $100~\text{s}^{-1}$. As this is three orders of magnitude larger than the inverse timescale at which the density of the condensate changes, we can assume that the BEC is in equilibrium throughout the excitation. 

As the interaction time is on the order of seconds, the condensate also decreases due to one-body collisions with the background gas. This is typically characterized by a loss rate $\Gamma$, and the simple differential equation 
\begin{align}
\frac{dN}{dt} = - \Gamma N
\end{align}
actually leads to a second term in the differential for the chemical potential as
\begin{align}
\frac{d\mu}{dt} = - \frac{2}{5} \Gamma \mu.
\end{align}
Two- and three-body losses can, in principle, be implemented in a similar fashion.

\subsection{Adding homogeneous broadening mechanisms}
We are not in the limit where the spectral linewidth of the system ($\sim$ few kHz) is negligible compared to the observed spectral feature ($\sim$ 15 kHz). In order to properly include a homogeneous broadening mechanism, we convolve the initial line shape with a broadening function $g(\nu)$ as
\begin{align}
\tilde{S}(\nu) = \int_{\nu'} \text{d} \nu' g(\nu-\nu') S(\nu').
\end{align}
This means that we do not have to change any of the first principle considerations, but can simply convolve our analytical result of the line shape with any broadening mechanism (which has to be independent of the spatial distribution of the atoms) without changing the formalism.

\subsection{Analytical result for $\tilde{S}(\nu)$}
Defining the Lorentzian distribution as
\begin{align}
g(\nu) = \frac{1}{2 \pi} \frac{\gamma}{\nu^2 + (\gamma/2)^2},
\end{align}
with $\gamma$ the full-width-half-max (FWHM), we obtain an analytical solution of the convolution with the line shape as defined in Eq. \ref{eqn:acstarklineshape}. We introduce (rescaled) variables to simplify the final expression:
\begin{align}
\tilde{\mu} &\equiv \Big( \frac{U_2}{U_1} - \frac{\alpha_2}{\alpha_1} \Big) \mu, \\
h\tilde{\nu} &\equiv h(\nu - \nu_0) + \Big( \frac{\alpha_2}{\alpha_1} - 1 \Big) (U_0 - \mu),\\
h \Delta &\equiv \sqrt{(h \tilde{\nu} - \tilde{\mu})^2 + (h \gamma/2)^2}, \\
h \delta_{\pm} &\equiv h \Delta \pm (h \tilde{\nu} - \tilde{\mu}), \\
h \xi_{\pm} &\equiv \sqrt{(h \gamma)^2 \pm 8 h \delta_{\pm} (h \tilde{\nu} - \tilde{\mu})}.
\end{align}
The line shape then becomes
\begin{widetext}
	\begin{multline}
			\tilde{S}(\nu) = \frac{30}{32} \frac{\hbar \Omega_R^2 N}{\tilde{\mu}^{\frac{5}{2}}} \Bigg[ 2 h \gamma \sqrt{\tilde{\mu}}\\
	+ \Bigg( h \tilde{\nu} \sqrt{h \xi_-} + \frac{h \gamma}{2} \sqrt{h \xi_+} \Bigg) \Bigg( \tan^{-1}\Bigg[ \frac{ - \sqrt{2 \delta_+ \tilde{\mu}/h}}{2 \Delta - \sqrt{2 \delta_- \tilde{\mu}/h}} \Bigg] - \tan^{-1}\Bigg[ \frac{ \sqrt{2 \delta_+ \tilde{\mu}/h}}{2 \Delta + \sqrt{2 \delta_- \tilde{\mu}/h}} \Bigg] \Bigg) \\
	- \Bigg( h \tilde{\nu} \sqrt{h \xi_+} - \frac{h \gamma}{2} \sqrt{h \xi_-} \Bigg) \Bigg( \text{Log} \Bigg[ \sqrt{\frac{h \delta_+ \tilde{\mu}}{2 (h \Delta)^2} + \Bigg( 1 - \sqrt{\frac{h \delta_- \tilde{\mu}}{2 (h \Delta)^2}} \Bigg)} \Bigg]
	-  \text{Log} \Bigg[ \sqrt{\frac{h \delta_+ \tilde{\mu}}{2 (h \Delta)^2} + \Bigg( 1 + \sqrt{\frac{h \delta_- \tilde{\mu}}{2 (h \Delta)^2}} \Bigg)} \Bigg] \Bigg) \Bigg].
	\end{multline}
\end{widetext}	

This expression is used in Eq. \ref{eqn:diffequation} to calculate the time-dependent line shape used to fit the data as shown in the paper.

\section{Rabi frequency}
Calculation of the Rabi frequency, which is an important parameter in our line shape calculations, requires trigonometry to calculate the projection of the spectroscopy beam on the quantization axis of the atoms. In the measurements used to determine the s-wave scattering length only the background magnetic field of the experiment was used. Here we briefly discuss how this field is measured and how we calculate the Rabi frequency from this.

We define the basic coordinate system where the $z$-axis is aligned with the axial direction (longitudinal axis) of the ODT and therefore of the BEC. In our experiment we have a set of coils in Helmholtz configuration (`fine tune coils') with their longitudinal axis parallel to this $z-$axis to apply a well-defined magnetic field if necessary.  As the quantization axis of the atoms is given by the magnetic field, we can calculate the magnetic field in this coordinate system and use it to calculate the projection of the spectroscopy beam polarization to calculate the Rabi frequency.

Using rf spectroscopy to measure the magnitude of the magnetic field, as used in Refs. \cite{Rooij1,Notermans1}, the fine tune coils are used to scan the magnetic field. The fine tune coils only produce a magnetic field component along the $z-$axis, and therefore these measurements can be used to determine the azimuthal angle $\theta = 52.1(4)^{\circ}$ of the background magnetic field with respect to the $z-$axis. With this angle we can calculate the projection $P$ of the polarization of the spectroscopy light on the quantization axis. To be more precise, we are interested in the orthogonal part of the polarization $P' = \sqrt{1 - P^2}$ as this contributes to the $\sigma^{-}$ transition that is required to induce the $2^3S_1 \ m_J = +1 \to 2^1S_0 \ m_J = 0$ transition. As the polar angle $\phi$ cannot be determined in the current setup, we take the full range as a conservative estimate to be $P^{'2} = 0.7 \pm 0.3$.

Using this calculated range, we can provide an estimate of the Rabi frequency $\Omega_R$ which is defined as
\begin{align}
\Omega_R^2 = \frac{6 \pi c^2}{\hbar \omega_0^3} A_{21} \vert \langle J_1 M_1 \vert q \vert J_2 M_2 \rangle \vert^2 I_0.
\end{align}
Here $\omega_0$ is the transition frequency, $A_{21} = 9.1 \times 10^{-8} \ \text{s}^{-1}$ the Einstein coefficient \cite{Rooij1}, $I_0 = 2P_0/\pi w_0^2$ the peak intensity with $P_0$ the total power and $w_0 = 0.3 \ \text{mm}$ the beam waist of the spectroscopy beam. The remaining matrix element $\vert \langle J_1 M_1 \vert q \vert J_2 M_2 \rangle \vert^2 = C_{C-G}^2 \cdot P'^2$, where $q = -1,0,+1$ represents the possible transitions (i.e. $\sigma^-, \pi, \sigma^+$, respectively), $C_{C-G}^2 = 1/3$ the corresponding Clebsch-Gordan coefficient of the transition, and $P'^2$ as discussed before. An additional factor $1/2$ needs to be included (i.e. $P'^2 \to P'^2/2$) to account for the fact that the orthogonal component of the linear polarization decomposes equally into left-hand and right-hand circular polarized light.

The final Rabi frequency estimate is 
\begin{align}
\Omega_R &= 2\pi \times (21 \pm 5) \ \text{Hz},
\end{align}
where the uncertainty is fully dominated by the determination of $P'^2$, and limits the systematic uncertainty in the determination of the s-wave scattering length.

We have also looked at possible elliptic polarization of our spectroscopy light. This effect is present, but the systematic effect on the Rabi frequency is smaller than the uncertainty in $P'^2$ and therefore not taken into the analysis for the s-wave scattering length.

\end{document}